\documentclass[10pt, technotes, twocolumn,twoside]{IEEEtran}
\IEEEoverridecommandlockouts
\usepackage{amsmath,amssymb,epsfig,cite,footnote,authblk}
\usepackage{tikz}
\usepackage{textcomp}
\usepackage{hyperref}

\usepackage{color}
\usepackage{hyperref}
\hypersetup{
    colorlinks,%
    citecolor=blue,%
    filecolor=blue,%
    linkcolor=blue,%
    urlcolor=black
}

\usepackage{changes}

\definecolor{bluecolor}{rgb}{0,0.,1.}

\definecolor{redcolor}{rgb}{.7,0.,0.}

\newcommand{\rev}[1]{{\color{black} #1}}

\definechangesauthor[name={Pratt}, color=red]{p}
\setremarkmarkup{(#2)}

\newcommand{\es}[1]{\begin{equation}\begin{split}#1\end{split}\end{equation}}

\newcommand{\V}{\mathcal{V}}

\newcommand{\br}{\boldsymbol{r}}
\newcommand{\bt}{\boldsymbol{t}}

\newcommand{\dd}{\textrm{d}}

\newcommand{\Ck}{\color{black}}

\newcommand{\squeezeup}{\vspace{-2mm}}
\newcommand{\squeezeupp}{\vspace{-3mm}}
\newcommand{\squeezeupsmall}{\vspace{-1mm}}
\newcommand{\fsize}{\fontsize{12pt}{6pt}\selectfont}
\newcommand\copyrighttext{%
  \fsize\textcopyright 2017 IEEE. Personal use is permitted, but republication/redistribution requires IEEE permission.
See\\ \url{http://www.ieee.org/publications_standards/publications/rights/index.html} for more information.
}
\newcommand\copyrightnotice{%
\begin{tikzpicture}[remember picture,overlay]
\node[anchor=south,yshift=750pt] at (current page.south) {\fbox{\parbox{\dimexpr\textwidth-\fboxsep-\fboxrule\relax}{\copyrighttext}}};
\end{tikzpicture}%
}

\graphicspath{{Pictures/}}%computer
\usepackage{epstopdf}
\title{\rev{Optimal} Non-uniform Deployments in Ultra-Dense Finite-Area Cellular Networks}
\author[1]{Pete Pratt}
\author[1]{Carl P. Dettmann}
\author[2]{Orestis Georgiou}
\affil[1]{School of Mathematics, University of Bristol, University Walk, Bristol, BS8 1TW, UK}
\affil[2]{Toshiba Telecommunications Research Laboratory, 32 Queens Square, Bristol, BS1 4ND, UK}
\begin{document}
\maketitle
\copyrightnotice
\thispagestyle{empty}
\begin{abstract}
Network densification and heterogenisation through the deployment of small cellular access points (picocells and femtocells) are seen as key mechanisms in handling the exponential increase in cellular data traffic. 
Modelling such networks by leveraging tools from Stochastic Geometry has proven particularly useful in understanding the fundamental limits imposed on network coverage and capacity by co-channel interference.
Most of these works however assume infinite sized and uniformly distributed networks on the Euclidean plane.
In contrast, we study finite sized non-uniformly distributed networks, and find the optimal non-uniform distribution of access points which maximises network coverage for a given non-uniform distribution of mobile users, and \textit{vice versa}.
%In this letter, we study finite sized non-uniformly distributed networks, and show that network performance is significantly different, and in fact, position dependent.
\end{abstract}
\squeezeup

\squeezeup\section{Introduction}\squeezeup

The rapid growth in cellular data traffic presents unprecedented strain on current cellular infrastructures.
One technology that can alleviate this is extreme network densification of cellular access points (APs) and also the heterogenisation of the conventional macrocell architecture with smaller picocells and femtocells.
Through the efficient exploitation of spatial frequency reuse, such dense heterogeneous cellular networks (HetNets) are expected to deliver  higher data rates as well as the ubiquitous wireless coverage needed in this internet age.
Therefore, network performance under spatial densification is an area of active academic and industrial research and standardisation with the view that the deployment of a multi-tier architecture can help facilitate the transition into 5G \cite{andrews2014will}.

The modelling and analysis of HetNets has been greatly facilitated by the development of Stochastic Geometry tools \cite{haenggi_book}, and through the establishment of meaningful performance metrics such as capacity, throughput, spectral efficiency, and coverage \cite{di2013average,elsawy2013stochastic}.
The main outcome of these research efforts has been the unveiling of engineering insights and a tractable framework for tackling network optimisation, e.g., designing Coordinated Multipoint (CoMP) transmission schemes \cite{nigam2014coordinated}.
As this research area matures however, some insights have been revisited and new complexities uncovered.
For instance, the celebrated result of \cite{andrews2011tractable} that coverage does not depend on network density when thermal noise is negligible has been overturned (by the same author) when considering multi-slope path loss models \cite{zhang2015downlink}.
A similar result was obtained by \cite{banani2015analyzing} who studied the impact  of AP density in finite-area networks.
\begin{figure}
  \centering
  \begin{minipage}[t]{0.22\textwidth}
    \includegraphics[width=\textwidth, trim = 0cm 1cm 1.3cm 0cm]{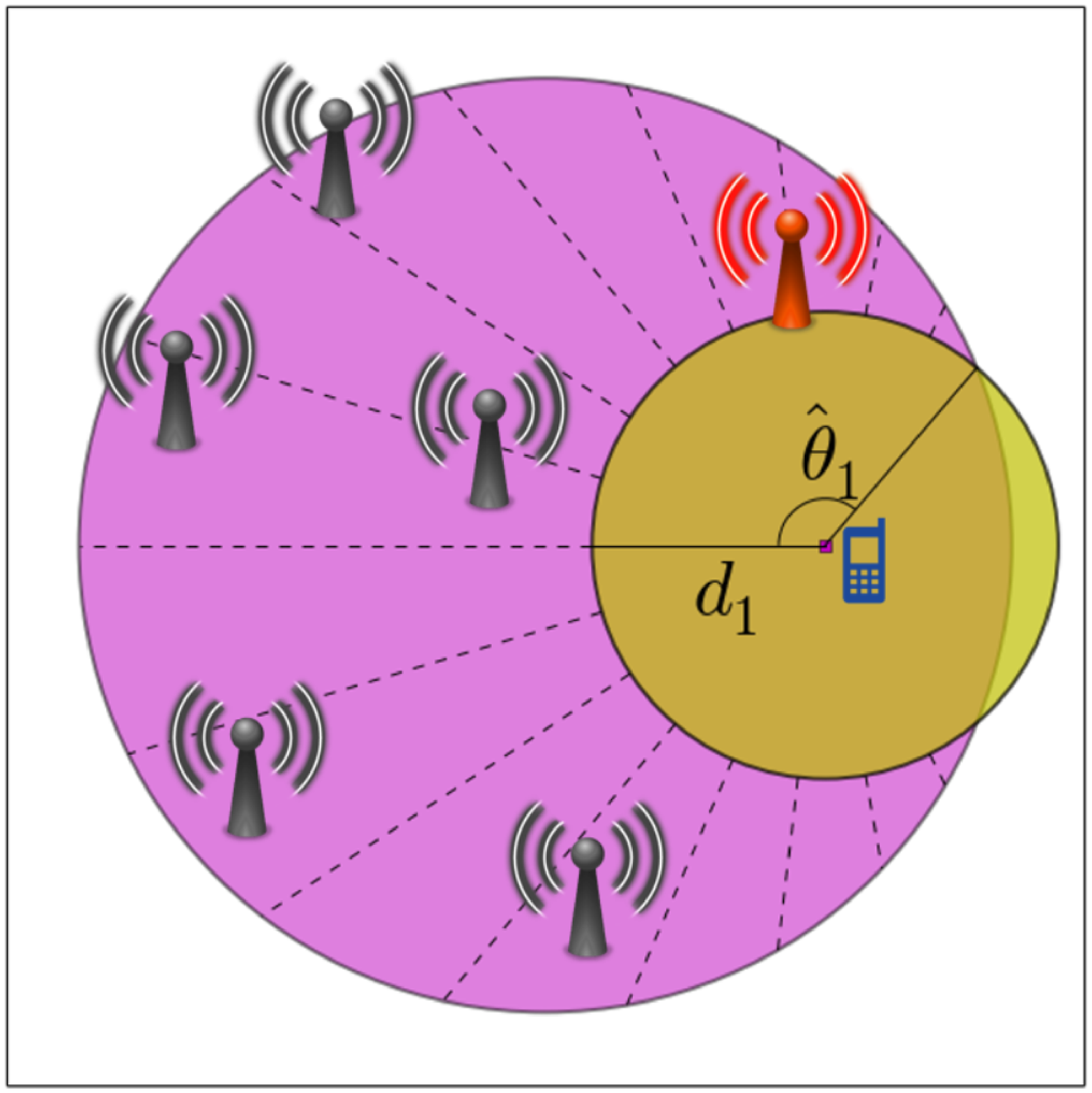}
  \end{minipage}
  \hfill
  \begin{minipage}[t]{0.22\textwidth}
	\includegraphics[width=\textwidth, trim = 1.3cm 1cm 0cm 0cm]{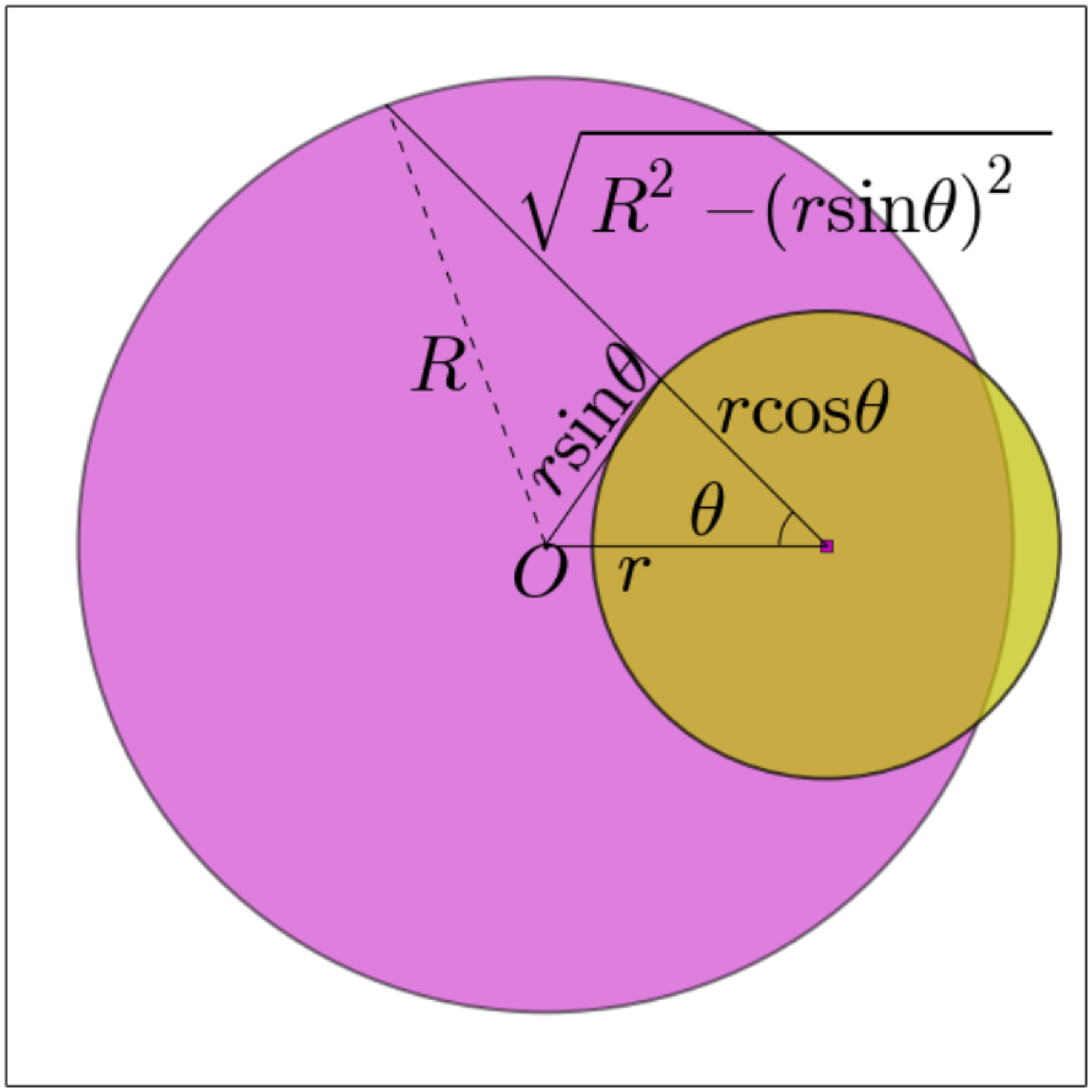}
\end{minipage}
\caption{\rev{Left: Simplified schematic of the system set up. A MU is located at $\br$, and is served by its nearest AP (the red one) which is located a distance $d_1$ away. Other APs are randomly distributed in the finite deployment region $\V$ according to a non-uniform density $\lambda(\bt)$, and act as sources of interference.  
Right: A schematic for the shift in coordinate system used to calculate \eqref{eq:Laplace}.}}
\label{fig:fig1}
\end{figure}
\Ck{
It is therefore desirable to understand how Access Points (AP) should be deployed in order to maximise Mobile User (MU) coverage; to this end, and motivated  by the aforementioned findings in \cite{nigam2014coordinated,andrews2011tractable,zhang2015downlink,banani2015analyzing}, we revisit the coverage problem in dense, finite area, cellular networks and analyse the optimal deployment of APs for a non-uniform distribution of MU. }\Ck{}

%The main aim of this paper is to analyse the optimal deployment of AP in ultra dense, finite networks for a general non-uniform distribution of MU.
We model the AP locations using a non-uniform Poisson Point Process (PPP) in a circular deployment region, and, assuming a closest AP user association model (as usually done when modelling cellular networks \cite{andrews2011tractable}), and derive novel analytic expressions for the probability density function (pdf) of the nearest neighbour distribution (NND). 
By leveraging tools from stochastic geometry, and  assuming Rayleigh fading, we calculate the position dependent interference field and outage probability, and use the NND to calculate \rev{a position dependent} coverage probability, i.e., the probability that a mobile user (MU) located at \rev{$\br\!\in\!\V$} can achieve a Signal-to-Interference-plus-Noise-Ratio (SINR) in the downlink greater than a threshold $q$.
Finally, we optimise the average coverage probability with respect to non-uniform AP and MU  spatial distributions.
The main contributions of this letter are:
\begin{itemize}
%\item an analysis of MU coverage (in the downlink) in an interference limited, finite network for a non-uniform AP deployment; 
%\item we analyse the optimal AP deployment (or equivalently random access transmission) scheme for ultra-dense networks with non-uniform distributions of MU.
%\item A semi-analytic expression for the connection probability following the SINR connection model assuming a nearest neighbour association model. 
%\item We numerically calculate the coverage probability (in the downlink) for an interference limited, finite network with a non-uniform AP deployment and highlight the interplay between border effects and AP distribution.  
%\item we provide an analytic expression for the NND in a finite domain for a non-uniform distribution of AP; %borders?
\Ck{}\item we analyse network coverage in a finite domain with non-uniform AP deployment for the first time, and highlight how border effects improve network coverage for convex AP deployments whilst the converse is true for the concave case. 
%the interplay between border effects and AP distribution. 
\item  we study the optimal AP deployment in finite regions which maximizes network coverage and show that for non-uniform MU spatial distributions the optimal AP deployment is non-trivial.
%\item we show that for convex MU distributions, the optimal AP deployment is convex, whilst for the concave case, for lower densities/path loss exponent or "less MU concavity'' a convex AP deployment is optimal.
%Namely, there is not a one-to-one matching of the MU distribution and the optimal AP deployment in finite domains.}\Ck{}
\end{itemize}
\Ck{
These results provide insight into how AP should be deployed or operated by highlighting the impact of non-uniform MU distributions, along with border and interference effects.
Note that even though we discuss a simplified non-uniform AP/MU distribution within a circular domain, our analysis provides insight for more complex AP/MU distributions and domains.}\Ck{}
\Ck{We  proceed by formally defining the system model, along with the NND and the connection model used in our main analysis in Section \ref{III}.}\Ck{}

\squeezeup\section{System Model}\squeezeup

\subsection{Network Model}\squeezeupsmall
We consider a non-uniform PPP of density $\lambda(\bt)\!=\!(t,\phi)$ in a circular disk $\mathcal{V} \subset \mathbb{R}^2$ of area $|\V|= \pi R^2$ and zero intensity elsewhere. 
Each point in our process corresponds to a cellular AP, transmitting at constant power $\mathcal{P}$.
For simplicity we restrict the analysis to quadratic radially symmetric AP distributions in polar coordinates such that the intensity function is  
\es{
\lambda(t) &= \lambda_0\left(a + bt^2\right),\:\:\ 0 \le \rev{t} \le R
\label{eq:Pdf}}
where $t$ is the radial distance, $\lambda_0>0$, $a\!=\!1\!-\! b R^2/2$ and $b\!\in\![-2/R^2, 2/R^2]$, such that there are, on average, $2\pi\int_0^R \lambda (\bt) t \dd t \!=\! \lambda_0 |\V|$ APs in $\V$. 
More specifically, the parameter $b$ in \eqref{eq:Pdf} controls how the AP are deployed within $\mathcal{V}$, and allows us to interpolate between three different AP deployment distributions, in particular: uniform $(b=0)$, convex ($b>0$ nodes located predominately near the border), and concave ($b<0$ nodes located near the centre of the deployment region). 
Note that the concave distribution is akin to the Random Waypoint Mobility model\footnote{In the RWPM nodes are initially placed in $\mathcal{V}$ according to some point process, where they then independently travel from waypoint to waypoint in a sequential manner. 
A node selects its next waypoint from a uniform distribution in $\V$, travels toward it in a straight line at a constant speed chosen at random and pauses with a certain probability, and then repeats the process.
This model gives rise to a stationary distribution with a higher density of nodes in the bulk of the domain due to the continual crossing of paths. 
The parameter $b \!=\! -2/R^2$ in \eqref{eq:Pdf} gives the RWPM in a disk with no pause time.} (RWPM)\cite{bettstetter2002spatial,gong2014interference,pratt} which models mobile ad hoc networks.
\rev{Note that by independently thinning a uniform PPP, one can easily obtain a non-uniform random access transmission scheme. 
The corresponding intensity would be $\lambda(t)\!\to\!  \lambda_0 p(t)$ where $p\!\in\![0,1]$ is a position dependent thinning probability \cite{haenggi_book}.}

Let the Euclidean distance from a receiving MU located at $\br \!=\! (r,\theta) \!\in\! \mathcal{V}$ and its distance to every other AP $i$ located at $\bt_i\!\in\!\V$ in the PPP be denoted by $d_i\!=\!|\br - \bt_i|$, where we order the distances as $0\!\leq\! d_1 \!\le\! d_2 \!\le\! ...$.
The assumption that a MU is served by its nearest AP is fairly intuitive since it is likely that the AP with the strongest signal will typically be the closest AP \cite{andrews2011tractable} (see Fig. \ref{fig:fig1}).
\squeezeup
\subsection{Nearest Neighbour distribution} \label{NND}
%------------------------------------------------------------------------------
% 								NND
%------------------------------------------------------------------------------
\squeezeupsmall
\begin{figure}
\centering
\includegraphics[width=0.4\textwidth, trim = 1cm 1cm 0cm 0cm]{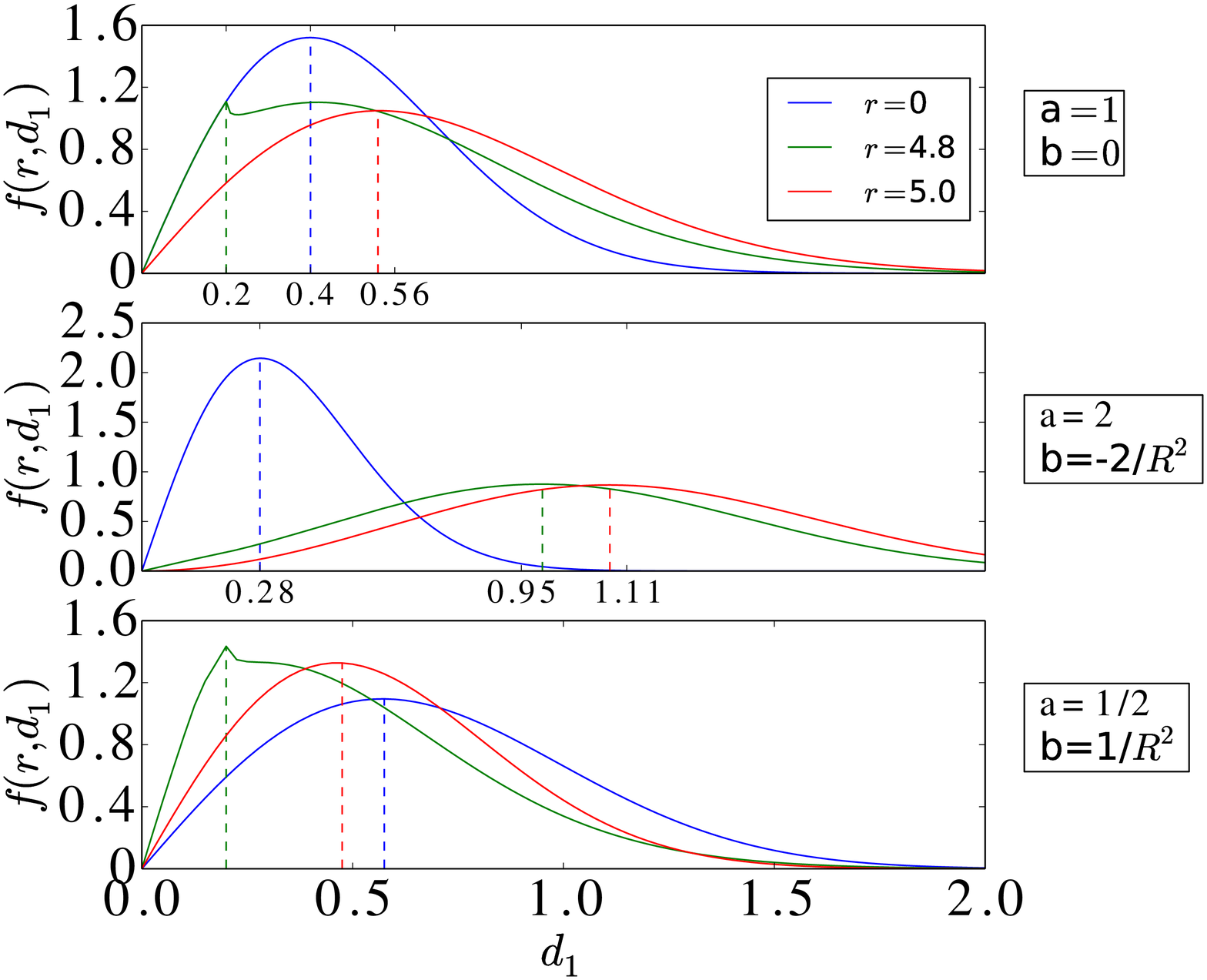}
\caption{Plot of the nearest neighbour pdf \eqref{eq:f} for MUs at different off-centre distances $r\!\in\![0,R]$ for different AP deployments: Uniform(top panel), Concave (middle), and Convex (bottom). We assume that $R\!=\!5$.}
\label{fig:pdf}
\end{figure}
In our model where a MU connects to its nearest AP, the pdf of this NND $f(\br,d_1)$  is given by the derivative of the contact distribution, or equivalently, minus the derivative of the void probability:
\es{f(\br,d_1) &= -\frac{\dd}{\dd d_1} \mathbb{P}\left[N(B(\br,d_1))=0\right]
\label{eq:VoidProb}}
Intuitively, the pdf of the NND in \eqref{eq:VoidProb} tells us the probability that the nearest AP will be a distance $d_1$ from a receiver located at $\br$.
Assuming the AP are distributed according to  \eqref{eq:Pdf}, the pdf of the NND in  \eqref{eq:VoidProb} becomes the radially symmetric and piecewise continuous function:
\es{f(r,d_1) &= \begin{cases}
f_1(r,d_1)\:\:\text{ if } r \le R - d_1\\%B(\br,d_1) \subset \mathcal{V}\\
f_2(r,d_1)\:\:\text{ if } r > R - d_1\end{cases}%B(\br,d_1) \cap \mathcal{V} \end{cases}
\label{eq:f}
}
\es{f_{1}(r,d_1)\!&=\!2\pi\!d_1\lambda_0\!\left(a\!+\!b(d_1^2\!+\!r^2)\right) e^{ -\lambda_0\pi d_1^2 \bigl(\!a\!+\!b\frac{d_1^2\!+\!2r^2}{2}\bigl) }
\label{eq:f1}} 
\es{\label{eq:f2}
&f_2(r,d_1) =  -\frac{\dd}{\dd d_1} \exp \Big[-2\lambda_0\Big\{a\tilde{r}\sqrt{d_1^2 - \tilde{r}^2}\\
&+d_1^2\Big(\arctan \Big[\frac{\tilde{r}}{\sqrt{d_1^2 - \tilde{r}^2}}\Big] + \frac{\pi}{2}\Big)\Big(a + \frac{b(d_1^2 + 2r^2)}{2}\Big) \\
&+ \frac{b\tilde{r}}{12\sqrt{d_1^2 - \tilde{r}^2}}(5d_1^4 - 7d_1^2\tilde{r}^2 + 2\tilde{r}^4) \\
&- \frac{b\sqrt{d_1^2 - \tilde{r}^2}}{12}\Big(d_1^2(13r + 3\hat{r}) + 2(r^2 + r^2\hat{r} +r\hat{r}^2 - 3\hat{r}^3)\Big) \\
&+ R^2\Big(a+\frac{bR^2}{2}\Big)\Big(\frac{\pi}{2} + \arctan \Big[\frac{\hat{r}}{\sqrt{R^2 - \hat{r}^2}}\Big]\Big) \\
&- \hat{r}\sqrt{R^2-\hat{r}^2}\Big(a + \frac{b}{6}(R^2 + 2\hat{r}^2)\Big)\Big\}\Big]
}
where we defined $\hat{r} \!=\! \frac{R^2 +r^2 - d_1^2}{2 r}$ and $\tilde{r} \!=\! \hat{r} \!-\! r$.

Fig.~\ref{fig:pdf} highlights the piecewise nature of $f(r,d_1)$, for the uniform (top panel), concave (middle panel) and convex (bottom panel) cases.
If the receiver is off centre, as the ball of radius $d_1$ grows, it will eventually intersect the boundary of the domain giving rise to this piecewise nature. 
%If we choose $0< \br  < R$ then we can choose $d_1$ such that the ball $B(\br,d_1) \subset \mathcal{V}$ but as we increase the ball size $d_1$ it intersects the boundary, which gives rise to this piecewise nature. 
%\es{\mathbb{E}[d_1] &= } 
\rev{This has a direct effect on the mean separation distance to the serving AP, given by $\mathbb{E}[d_1]\!=\!\int_0^{d_{\text{max}}}f(r,d_1)d_1\dd d_1$ which clearly depends on $r$ and $b$.
For uniform deployments, $\mathbb{E}[d_1]$ is smaller for a MU located near the centre ($r\!\approx\!0$), and bigger near the border ($r\!\approx\!R$).
This border effect is amplified in the concave case, and reversed in the convex AP deployment density.}

%------------------------------------------------------
%				Connection Model
%------------------------------------------------------
\squeezeup\squeezeup
\subsection{Connection Model}\label{CM}\squeezeupsmall

%\Ck{The attenuation in the wireless channel is modelled as a product of the large scale path loss component and small scale fading component. }\Ck{}
The signal power received by a MU in the far field is inversely proportional to the separation distance \rev{$x$} between source and destination.
We adopt the following well-established pathloss attenuation function $g(x) \!=\! x^{-\eta}$,
where $\eta\!\geq\! 2$ is the pathloss exponent.
For free space $\eta \!=\! 2$ so the signal strength obeys exactly the inverse square law, and decays faster for more cluttered environments; $\eta \!=\! 4$ is typically taken for urban areas. 
In addition to pathloss attenuation, small-scale fading can affect the received signal power. 
%The latter models the fast time variations in the channel gain due to the transmitted signal being reflected and scattered multiple times resulting in constructive and destructive interference at the receiver. 
We model this by a Rayleigh fading channel, such that the gain between a transmitting AP $i$ and the receiving MU is an independent random variable with a standard exponential distribution $|h_i|^2\! \sim\! \exp(1)$, for $i=1,2,\ldots$ 
%Multi-slope non-singular pathloss functions and generalised fading distributions (including shadowing) have been studied in \cite{nguyen2016coverage}.
%We do not consider any large-scale fading effects.
We therefore formulate the received SINR at the MU located at $\br$ in the downlink as 
$\text{SINR}\!=\! \frac{\mathcal{P} |h_1|^2 g(d_1)}{\mathcal{N}+ \mathcal{I}}$ where $\mathcal{I} \!=\! \mathcal{P}\sum_{k \geq 2}g(d_k)|h_k|^2$ is the aggregate interference from all other transmitters, $\mathcal{P}$ is the transmit power (assumed the same for all AP), and $\mathcal{N}$ is the average thermal noise power.
For the sake of simplicity, we consider the quality of the received signal to be completely characterised by the SINR.

We assume that a MU can connect to its nearest AP if the SINR at the MU is greater than a threshold $q$, else it is said to be in outage. 
The connection probability is thus given by, $H_1(\br,d_1) \!=\mathbb{P}[\text{SINR} \ge q$].

\begin{figure}
\centering
\includegraphics[width=0.47\textwidth, trim = 0cm 2cm 0cm 0cm]{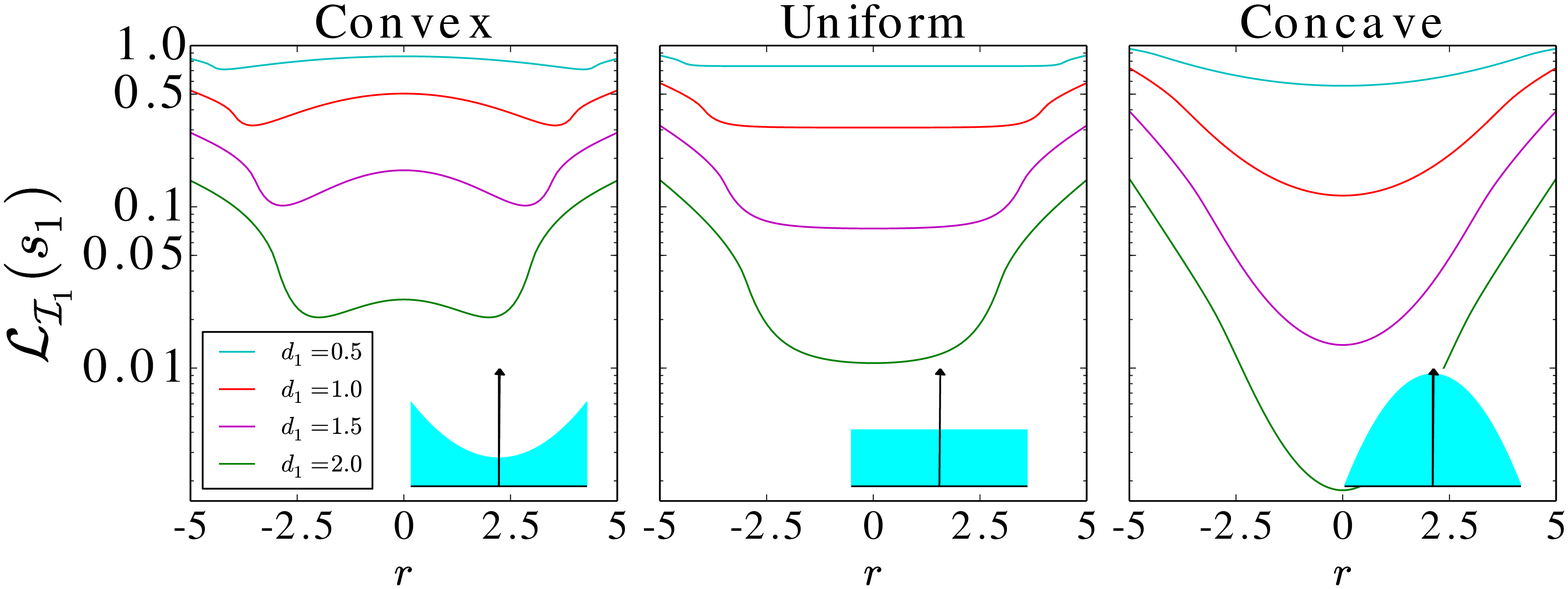}
\caption{The Laplace functional \eqref{eq:Laplace2} plotted as a function of MU position $r$ for all three cases with the distributions inset.
Parameters used: $\eta = 6$, $\mathcal{P}, q =1, \lambda_0 = 1$. Note: For $\eta < 6$ we have a decrease in $\mathcal{L}_{\mathcal{I}_1}(q \,d_1^\eta)$.}
\label{fig:IntFig}
\end{figure}

Conditioning on the interference we have that the connection probability is given by
\es{H_1(\br,d_1) &=\mathbb{E}_{\mathcal{I}}\Big[\mathbb{P}\Big[|h_1|^2 \ge \frac{q(\mathcal{N} + \mathcal{I})}{g(d_1)\mathcal{P}}\Big|\br, d_1, \mathcal{I}\Big]\Big]\\
%&=\mathbb{E}_{\mathcal{I}_1}\left[\exp\left[-\frac{q(\mathcal{N} +  \mathcal{I}_k)}{g(d_1)\mathcal{P}}\right]\right]\\
 &=\exp \Big[-\frac{q\mathcal{N}}{\mathcal{P} g(d_{1})}\Big] \mathcal{L}_{\mathcal{I}}(q \, d_1^\eta)\label{eq:H1b}}
where $\mathcal{L}_{\mathcal{I}}(s)$ is the Laplace transform of the random variable $\mathcal{I}$. 
Assuming that $|h_{k}|^2 \!\sim\! \exp(1)$ and evoking the probability generating function for a PPP \cite{haenggi_book} we can express this as
\es{
\mathcal{L}_{\mathcal{I}}(q \, d_1^\eta) = \exp\Big[-\int_{\mathcal{V}\backslash B(\br, d_1)}\frac{\lambda(z) }{1 + \frac{d_k^{\eta}}{q \, d_1^\eta }} d_k \dd d_k \dd \theta \Big]
\label{eq:Laplace}}
where $z\!=\!\sqrt{r^2 + d_k^2 - 2rd_k\cos\theta}$.
Notice that the integral is computed over the whole domain $\mathcal{V}$ but excluding the ball $B(\br, d_1)$, see Fig.~\ref{fig:fig1}. 
Moreover, the integral is over a non-uniform density $\lambda(z)$ where we have also used the cosine rule to shift the polar coordinate system from the centre of $\V$, to that centred on $\br$. 
This shift in coordinates allows us to further simplify the Laplace functional and arrive at 
\es{\mathcal{L}_{\mathcal{I}}(q \, d_1^\eta) &=\exp \biggl[-\lambda_0 \int_{0}^{\hat{\theta}_1} \phi(\hat{R}) - \phi(d_1)\dd \theta \biggl]
\label{eq:Laplace2},}
where, 
$\hat{\theta}_1 \!=\!  \min\Big[\arccos\Big(\frac{r^2 \!+\! d_1^2 \!-\! R^2}{2rd_1}\Big),\pi \Big]$
is the angle of intersection shown in Fig. \ref{fig:fig1},
$\hat{R} \!=\! r\cos\theta + \sqrt{R^2 - r^2 \sin^2 \theta}$ is the radial distance from the MU to the domain border, and 
\es{\phi(x) &= \frac{x^2}{6}\Big[6(a+br^2)\, \psi(\frac{2}{\eta},\frac{x^{\eta}}{q \, d_1^{\eta}}) \\
&+b x \Big(3x\;\psi(\frac{4}{\eta},\frac{x^{\eta}}{q \, d_1^{\eta}})
- 8 r\cos \theta\psi(\frac{3}{\eta},\frac{x^{\eta}}{q \, d_1^{\eta}})\Big)\Big]
\label{eq:L_General}}
is the result after calculating the radial integral over $d_k$ in \eqref{eq:Laplace}, where we also have defined $\psi(x,y) \!=\! {}_2F_1\big(1,x,1\!+\!x,-y\big)$.
Here ${}_2F_1$ is the Gauss hypergeometric function.
Equation \eqref{eq:Laplace2} can only be expressed in closed form for the special case of $\br \!=\! 0$, in which case we obtain
\es{\mathcal{L}_{\mathcal{I}}(q \,d_1^\eta)&=\exp\bigl[\frac{2 d_1^2q (2 + b d_1^2 - bR^2)}{4(1+q)}\\
&- 4R^2\psi(\frac{2}{\eta},\frac{R^{\eta}}{q \,d_1^{\eta}} )  + bR^4 \psi(\frac{4}{\eta},\frac{R^{\eta}}{q \,d_1^{\eta}}) \bigl],\label{eq:L_centre}}
but can be computed numerically using standard libraries.
Fig.~\ref{fig:IntFig} shows the calculation of \eqref{eq:Laplace2} as a function of radial position of the MU $r$ conditioning on different values of $d_1$.
Note how the interference is reduced, in all cases, when $r$ is near the border. 
This is because the expected separation distance of the interfering APs $\mathbb{E}[d_k]$, $k \!>\!1$, typically increases as a MU approaches the border, thus leading to a weaker interference field.
\rev{This is purely a geometrical effect. Intuitively, when $r\!=\!0$ all interferers are located at distances $d_{k}\!\in\![d_1,R]$.
In contrast, when $r\!=\!R$ all interferers are located at distances $d_{k}\!\in\![d_1,2R]$. }
Again, a non-uniform density of APs can amplify or counter this effect. 
Further, for areas of low density (e.g near the centre for the convex case), the interference field is also reduced. 
By increasing $d_1$, $\mathcal{L}_{\mathcal{I}_1}(q \,d_1^\eta)$ decreases, (interference increases); a consequence of $\frac{d_k}{d_1}$ (integrand of \eqref{eq:Laplace}) decaying to one.
In other words, conditioning on the AP being further away the ratio $\mathbb{E}[\frac{d_k}{d_1}] \to 1$ which in turn will cause the average Signal-to-Interference-Ratio (SIR) $\to 0$. 
\Ck{We proceed by using the aforementioned NND %given by eq \eqref{eq:f} 
and connection model %eq. \eqref{eq:Laplace} 
to analyse MU coverage and the optimal AP deployment. }\Ck{}
%------------------------------------------------------------------------------
% 								 Coverage
%------------------------------------------------------------------------------
\squeezeupp
\section{Optimal Coverage Deployment} \label{III}
\squeezeupsmall
\subsection{Mobile User Coverage}\squeezeup
An important performance metric commonly used is the network coverage probability experienced by a MU given by,
\es{
C(\br,b,\lambda_0\pi R^2) &= \int_{0}^{d_{\text{max}}} H(\br ,d_1) f(\br , d_1) \dd d_1\label{eq:Coverage}
}
where $H(\br ,d_1)$ is the connection probability given in \eqref{eq:L_General}, $d_{\text{max}}\!=\!R\!+\!r$ is the maximum distance from the MU to the boundary, and $f(\br , d_1)$ is the pdf of the NND given in eq.\eqref{eq:VoidProb}.

\begin{figure}
\centering
\includegraphics[width= 0.45\textwidth, trim = 2cm 1cm 1cm 0cm]{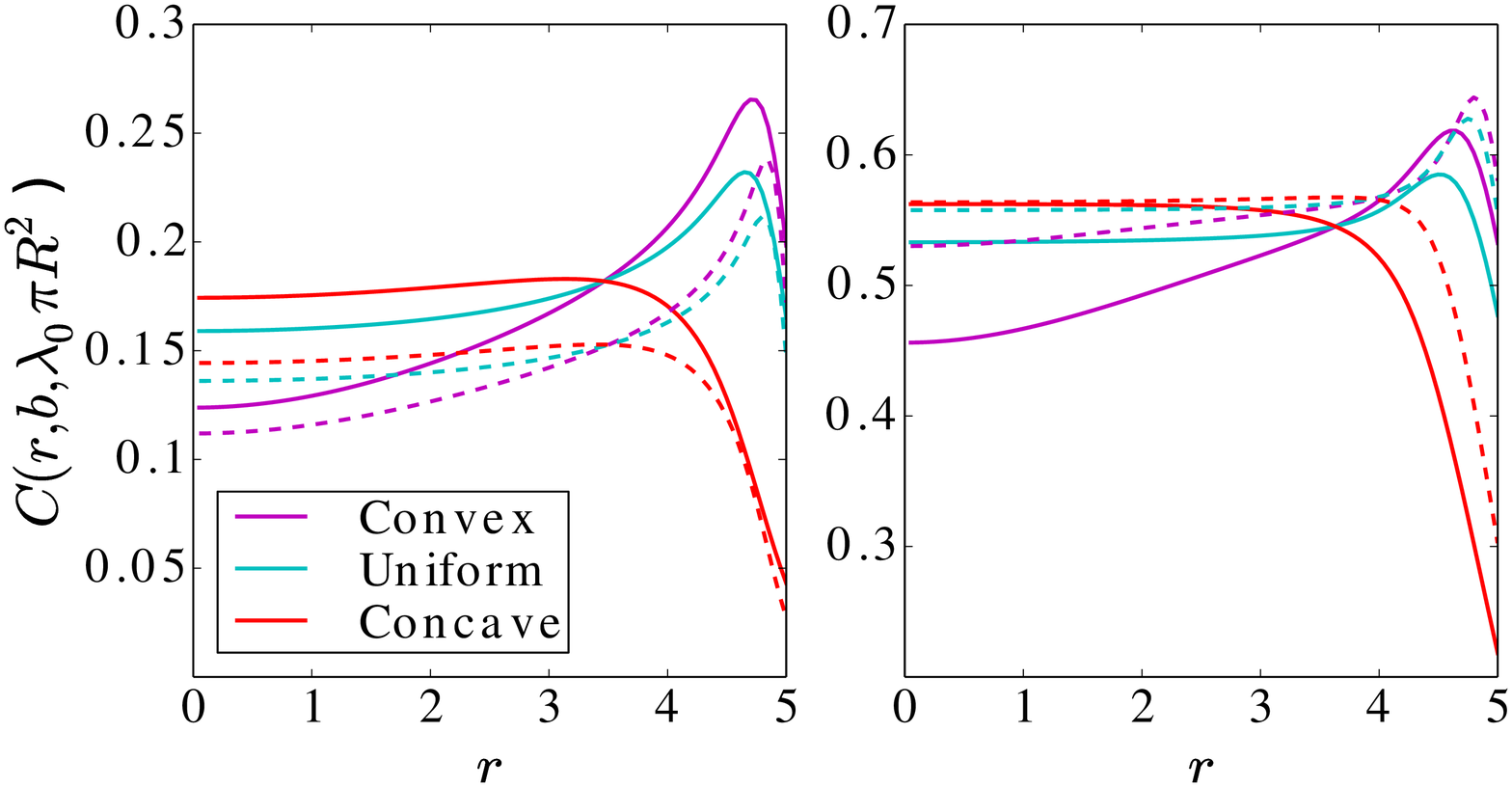}
\caption{The coverage probability, \eqref{eq:Coverage}, as a function of position $r$ from the centre to the boundary, with $\mathcal{P} = \mathcal{N} = q = 1$, $R = 5$ and $\eta = 2,4$ for the left and right panel respectively. Solid line: $\lambda_0 = 1$; dashed line: $\lambda_0 = 5$.}
\label{fig:CovFig}
\end{figure}

More specifically, eq \eqref{eq:Coverage} tells us the probability that a MU at $\br$ can successfully decode a message from its nearest AP, and this probability depends on both the distribution of APs,$\lambda(t)$ controlled by $b$, and the location of the MU, see Fig~\ref{fig:CovFig}.
In fact, since the AP are deployed in a radially symmetric fashion, following  eq \eqref{eq:Pdf}, the coverage probability is also radially symmetric as a result.
%\rev{Of course, the coverage is not only position dependent with respect to $r$, but also strongly dependent on the AP deployment distribution $\lambda(t)$ controlled by $b$ which we focus on next.}
\squeezeup
\subsection{Optimal Coverage Deployment}\squeezeupsmall
It is important to know how to deploy APs such that MU connectivity can be maximised.
%, or num. isolated nodes is minimised. 
To this end, we introduce the average coverage probability for a radially symmetric non-uniform distribution of MUs given by $\rho(r) \!=\! 1-\beta \frac{R^2}{2} + \beta r^2$, where $\beta$ plays a similar role as $b$ in \eqref{eq:Pdf} and  $\int_{\mathcal{V}} \rho(r) r \dd r \!=\! |\mathcal{V}|$:
\es{\bar{C}(b,\beta,\lambda_0\pi R^2) = \frac{2}{R^2} \int_0^R \rho(r) C(r,b,\lambda_0\pi R^2) r \dd r
\label{eq:AveCoverage}} 
Eq. \eqref{eq:AveCoverage} can be maximised to give the optimal value of $b$ 
\es{b^*(\eta,\beta,\lambda_0\pi R^2) = \arg\max_{b} \bar{C}(b,\beta,\lambda_0\pi R^2) \label{eq:b*},}
conditioned on $\eta, \beta$ and density of APs $\lambda_0$ in $\V$. 
\rev{Equation \eqref{eq:b*} is a maximization over a double integral of an exponential of another integral, and is therefore computationally taxing. The simplifications and analysis performed in the previous sections have alleviated this task to some extent.}
\squeezeupsmall
\squeezeupp\squeezeupsmall\subsection{Numerical Results}\squeezeupsmall

Fig~\ref{fig:CovFig} shows how the border effects result in a drop in coverage probability due to low $f(r,d_1)$,  and the distance between the first and second nearest neighbour is likely to be much less compared with a MU located near the centre of $\V$. 
Similarly, away from the border, areas of high AP density mean that coverage  decreases due to a higher interference field, regardless of the close proximity of the nearest neighbour. 
%As a result,  both the centre and border MU coverage suffers in the concave case.
For the convex case we observe a ``sweet spot'' where the trade-off between the distribution of APs and border effects helps to maximise coverage probability.

\begin{figure}
\centering
\includegraphics[width= 0.5\textwidth, trim = 3cm 2cm 3cm 0cm]{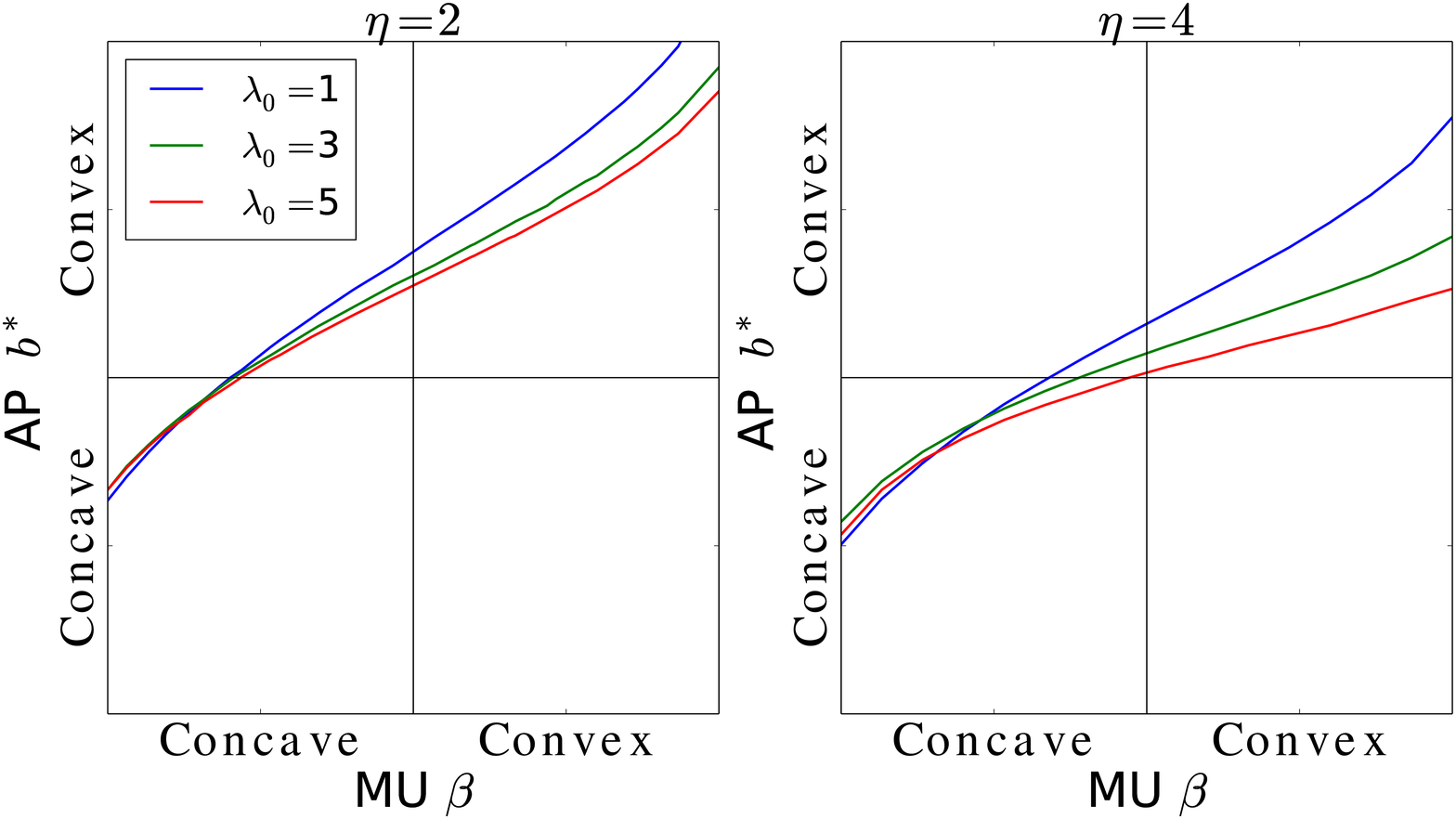}
\caption{The optimal distribution of AP given a deployment of MU for different $\lambda_0$ with $R= 5$, and $\eta = 2,4$ for the left and right panel respectively.}
\label{fig:OptFig}
\end{figure}

\rev{Fig.~\ref{fig:OptFig} shows the optimal deployment of APs $b^*$ as a function of the non-uniform MU distribution parameter $\beta$, for different values of $\eta$ and $\lambda_0$.
Basically, these plots show that a convex spatial distribution of MUs (representing the demand of data in the downlink) should be met by a convex spatial distribution of active APs (representing the supply of data in the downlink).
For a concave distribution of MUs in a finite domain however, the optimal AP distribution will  depend on $\beta, \eta,$ and $\lambda_0$, and may vary from concave to convex.
This is an interesting observation caused by the trade-off between interference and border effects, revealed for the first time in this paper.
%The AP distribution attempts to align itself more closely to that of the MU one in extremal regions (i.e $b = \pm 2/R^2$); however, it does not match it directly as this would lead to regions with a high interference field.
%The benefits of having an extremely concave AP distribution are minimal due to the additive nature of interference.   
Note that the optimal distribution of APs in ultra-dense deployments, i.e., for $\lambda_0\!\gg\!1$, tends to be more uniform.
A similar trend is observed for larger pathloss exponents $\eta \ge 2$.}
%\Ck{We note that by varying $\lambda_0$ it is equivalent to changing $R$ due to how the parameters scale.
%Furthermore, if we let $R\to \infty$, then the MU distribution becomes uniform thus the optimal AP deployment will also be uniform.}\Ck{}
Finally, recall that a uniform deployment of APs can be made non-uniform via a non-uniform independent thinning process $p(\bt)$ modelling a random access transmission scheme.

%-----------------------------------------------------------------------------------------------
%												Conclusion
%-----------------------------------------------------------------------------------------------
\squeezeupp\squeezeupsmall\section{Conclusion}\squeezeup
In this letter we analyse network densification, a key aspect in providing the increased network performance promised by the fifth generation of wireless communications. 
%Assuming a nearest neighbour association scheme we show that under the Signal-to-Interference-Plus-Noise-Ratio connection model that mobile users located near the borders and regions of high density will be least likely to connect. 
Under a nearest neighbour association scheme we highlight the impact that the spatial distributions of APs and MUs have on network coverage in finite operating regions.
Using tools from stochastic geometry, we investigate the trade-off in SINR and how this is affected by the position of MUs in a finite region, and also the spatial distribution of APs.
Finally, we formulate an optimisation problem for the average network coverage by adapting the APs deployment method or transmission scheme according to the spatial distribution of MUs. 
An application where an adaptive transmission scheme could be used to achieve optimal MU coverage would be for cities where the MUs spatial distribution goes from concave during work-hours, to convex at night-time.
%One possible extension is to study MU coverage with Collaborative AP transmission schemes where the second nearest AP (strongest interferer) collaborates with the nearest AP in a maximum ratio transmission (MRT). Alternatively, joint transmission scheme such as in CoMP or other distributed antenna systems could be studied in order to improve network performance. 
Collaborative AP transmission schemes where the $k$ nearest APs  collaborate in a maximum ratio transmission (MRT) or joint transmission scheme such as in CoMP could be studied in order to improve network performance.

\squeezeupp
\section*{Acknowledgements}\squeezeup
The authors would like to thank the directors of the Toshiba Telecommunications Research Laboratory for their support.
This work was supported by the EPSRC [grant number EP/N002458/1].
In addition, Pete Pratt is partially supported by an EPSRC Doctoral Training Account. 
\squeezeup\squeezeupsmall

%%%%%%%%%%%%%%%%%%%%%%

\bibliographystyle{ieeetr}
\bibliography{Cov_bib}

\begin{thebibliography}{10}

\bibitem{andrews2014will}
J.~G. Andrews, S.~Buzzi, W.~Choi, S.~V. Hanly, A.~Lozano, A.~C. Soong, and
  J.~C. Zhang, ``What will {5G} be?,'' {\em IEEE Journal on Selected Areas in
  Communications}, vol.~32, no.~6, pp.~1065--1082, 2014.

\bibitem{haenggi_book}
M.~Haenggi, {\em Stochastic geometry for wireless networks}.
\newblock Cambridge University Press, 2012.

\bibitem{di2013average}
M.~Di~Renzo, A.~Guidotti, and G.~E. Corazza, ``Average rate of downlink
  heterogeneous cellular networks over generalized fading channels: A
  stochastic geometry approach,'' {\em IEEE Transactions on Communications},
  vol.~61, no.~7, pp.~3050--3071, 2013.

\bibitem{elsawy2013stochastic}
H.~ElSawy, E.~Hossain, and M.~Haenggi, ``Stochastic geometry for modeling,
  analysis, and design of multi-tier and cognitive cellular wireless networks:
  A survey,'' {\em IEEE Communications Surveys \& Tutorials}, vol.~15, no.~3,
  pp.~996--1019, 2013.

\bibitem{nigam2014coordinated}
G.~Nigam, P.~Minero, and M.~Haenggi, ``Coordinated multipoint joint
  transmission in heterogeneous networks,'' {\em IEEE Transactions on
  Communications}, vol.~62, no.~11, pp.~4134--4146, 2014.

\bibitem{andrews2011tractable}
J.~G. Andrews, F.~Baccelli, and R.~K. Ganti, ``A tractable approach to coverage
  and rate in cellular networks,'' {\em IEEE Transactions on Communications},
  vol.~59, no.~11, pp.~3122--3134, 2011.

\bibitem{zhang2015downlink}
X.~Zhang and J.~G. Andrews, ``Downlink cellular network analysis with
  multi-slope path loss models,'' {\em IEEE Transactions on Communications},
  vol.~63, no.~5, pp.~1881--1894, 2015.

\bibitem{banani2015analyzing}
S.~A. Banani, A.~W. Eckford, and R.~S. Adve, ``Analyzing the impact of access
  point density on the performance of finite-area networks,'' {\em
  Communications, IEEE Transactions on}, vol.~63, no.~12, pp.~5143--5161, 2015.

\bibitem{bettstetter2002spatial}
C.~Bettstetter, C.~Wagner, {\em et~al.}, ``The spatial node distribution of the
  random waypoint mobility model.,'' {\em WMAN}, vol.~11, pp.~41--58, 2002.

\bibitem{gong2014interference}
Z.~Gong and M.~Haenggi, ``Interference and outage in mobile random networks:
  Expectation, distribution, and correlation,'' {\em Mobile Computing, IEEE
  Transactions on}, vol.~13, no.~2, pp.~337--349, 2014.

\bibitem{pratt}
P.~Pratt, C.~P. Dettmann, and O.~Georgiou, ``How does mobility affect the
  connectivity of interference-limited ad-hoc networks?,'' in {\em Modeling and
  Optimization in Mobile, Ad Hoc, and Wireless Networks (WiOpt), 2016 14th
  International Symposium on}, pp.~000--001, IEEE, 2016.

\end{thebibliography}

\end{document}